\hfuzz 2pt
\font\titlefont=cmbx10 scaled\magstep1
\magnification=\magstep1

\null
\vskip 1.5cm
\centerline{\titlefont PLANCK'S SCALE DISSIPATIVE EFFECTS}
\medskip
\centerline{\titlefont IN ATOM INTERFEROMETRY}
\vskip 2.5cm
\centerline{\bf F. Benatti}
\smallskip
\centerline{Dipartimento di Fisica Teorica, Universit\`a di Trieste}
\centerline{Strada Costiera 11, 34014 Trieste, Italy}
\centerline{and}
\centerline{Istituto Nazionale di Fisica Nucleare, Sezione di
Trieste}
\vskip 1cm
\centerline{\bf R. Floreanini}
\smallskip
\centerline{Istituto Nazionale di Fisica Nucleare, Sezione di
Trieste}
\centerline{Dipartimento di Fisica Teorica, Universit\`a di Trieste}
\centerline{Strada Costiera 11, 34014 Trieste, Italy}
\vskip 2cm
\centerline{\bf Abstract}
\smallskip
\midinsert
\narrower\narrower\noindent
Atom interferometers can be used to study phenomena leading
to irreversibility and dissipation, induced by the dynamics
of fundamental objects (strings and branes) at a large mass scale.
Using an effective, but physically consistent description in
terms of a master equation of Lindblad form, the modifications of
the interferometric pattern induced by the new phenomena
are analyzed in detail. We find that present experimental
devices can in principle provide stringent bounds on the
new effects.
\endinsert
\bigskip

\vfil\eject

{\bf 1. INTRODUCTION}
\medskip

The evolution in time of a system $\cal S$ immersed in a large
environment $\cal E$ can be obtained from the dynamics of the
total system $\cal S+E$ by eliminating ({\it i.e.} integrating
over) the degrees of freedom of $\cal E$. Since $\cal S+E$ is
closed, the total dynamics is unitary; this is no longer true
for the evolution of the subsystem $\cal S$ alone, which in general
turns out to be very involved, developing non-linearities and memory
effects. However, when the interaction between subsystem and
environment is weak and there are no initial correlations between
$\cal S$ and $\cal E$, the time evolution of $\cal S$ can still
be realized through linear maps on the states of $\cal S$,
satisfying basic physical requirements, like forward in time
composition law (semigroup property), entropy increase (irreversibility)
and complete positivity (that guarantees the physical consistency
of the evolution in all situations). These one-parameter (=time)
family of maps form a so called quantum dynamical semigroup [1-4]
and are generated by a master equation of Lindblad form [5].

This description of the time-evolution of open systems is very
general; it was originally developed in the framework of quantum
optics [6-8], and subsequently used to model very different physical
situations, from the study of various statistical systems [1-3],
to the analysis of the interaction of a microsystem with a
macroscopic measuring apparatus [9-11], to the description of the
emergence of the classical world [12, 13] and of the
so called dynamical reduction [14].

Master equation of Lindblad form can also be used to describe
phenomena leading to irreversibility and dissipation at low
energy generated by the dynamics of fundamental objects
at a large scale, typically the Planck mass [15]. Indeed, the
dynamics of extended objects, strings and branes, gives
rise at low energies to a weakly coupled heat bath, and as
a consequence to decoherence phenomena [16]. From a more
phenomenological point of view, similar effects have also
been described in the framework of quantum gravity [17]: due to
the quantum fluctuations of the gravitational field and
the possible generation of virtual black holes, spacetime
should become ``foamy'' at scales comparable to the Planck
length, leading to loss of quantum coherence [18-23].
Furthermore, dissipation and decoherence are also the natural
outcome of the general dynamics in theories with large extra
dimensions [24]: indeed, the possible energy leakage from the
boundary of spacetime (our four-dimensional brane universe)
into the bulk due to gravity effects would inevitably inject
noise into the boundary, thus inducing irreversibility and
dissipation at low energy.

Our present knowledge of string theory does not allow precise
estimates of the magnitude of these non-standard effects.
Nevertheless, dimensional arguments suggest that they must be
very small, being suppressed by at least one inverse power
of a large fundamental mass.
Despite of this, they can be in the reach of various
interferometric devices. Indeed, detailed investigations involving
different elementary particle systems (neutral mesons [25, 26],
neutrons [27], photons [28], and neutrinos [29])
have shown that present and future experiments might soon reach
the sensitivity required to detect the new, non-standard phenomena.

Another physical situation in which phenomena leading to
irreversibility and dissipation can be studied is provided by
atom interferometers, where a beam of nearly monoenergetic atoms 
is coherently split into two components that are recombined at the exit
of the apparatus [30-33].
The interferometric pattern observed at the end of the devise
is influenced by the action of
external phenomena, produced {\it e.g.} by an external electric
or magnetic field, or by earth gravity. The sophistication
of present interferometric apparata is so high that the
theoretically predicted changements in the interferometric
figure for some of these phenomena have been confirmed
with high accuracy [30-32].

Irreversibility and dissipation also affect the propagation
of the atoms in the interferometer; this leads to a deformation
of the corresponding interferometric pattern at the exit of the
apparatus. It turns out that these modifications are very
distinctive of the dissipative phenomena, and can not be mimicked
by other physical effects, as the ones mentioned before.

In the following, we shall analyze in detail these modifications
under the hypothesis that the generalized dynamics of the
atoms inside the interferometer be generated by a master equation
of Lindblad form. For sake of definiteness, we shall limit our discussion
to three-grating atom interferometers in the Bragg regime, where the
split and the recombination of the incident beam is realized by
material gratings or laser standing waves.
The approach has points in common with
the one adopted in [27] to study similar effects
in neutron interferometry [34].
The perturbative treatment adopted there is however
inapplicable in the present case: this fact, together
with the operational differences in the actual functioning
of an atom interferometer require a completely different
and independent analysis. As discussed in the final section, 
the outcome of our investigations 
is that atom interferometry experiments
could provide the most accurate estimate of the non-standard,
dissipative effects that can be induced by a fundamental
dynamics at Planck's scale.

\vskip 1cm

{\bf 2. MASTER EQUATION}
\medskip

The evolution of the atoms inside the interferometer can be
analyzed using an abstract, two-dimensional Hilbert space.
The states corresponding to the two split beams in the
apparatus can be taken to be the basis states in this space.
More in general, the quantum state of an atom travelling inside
the interferometer will be a statistical mixture of the
basis states, and therefore described by a density matrix
$\rho$, {\it i.e.} by a hermitian, positive defined operator
with unit trace.%
\footnote{$^\dagger$}{For earlier works on the use of the formalism
of density matrices in atom interferometry, see [35-37, 33]
and references therein.}
With respect to the chosen basis,
one can then write:
$$
\rho=\left(\matrix{
\rho_1 & \rho_3\cr
\rho_4 & \rho_2}\right)\ ,
\qquad  \rho_4=\rho_3^*\ ,
\qquad \rho_1+\rho_2=1\ ,
\eqno(2.1)
$$
where $*$ signifies complex conjugation.

As explained in the introductory remarks,
the starting point of our analysis is the assumption that
the dynamics of the atoms inside the interferometer be
generated by a master equation of Lindblad form%
\footnote{$^\ddagger$}{An equation of this type has also been used
to study decoherence effects in position space
induced by the scattering of photons on the atoms
inside the interferometer [37]. Instead, the evolution equation (2.2)
is written in ``polarization'' space, and, as explained below,
represents the most general master equation compatible with basic 
physical requirements.}
[1-5]:
$$
{\partial\over\partial t}\rho(t)=-i\big[H , \rho(t)\big]+
D[\rho(t)]\ .
\eqno(2.2)
$$
The first term in the r.h.s. represents the standard hamiltonian
contribution. In the chosen basis, the effective hamiltonian can
be written as:
$$
H=\left(\matrix{
E+\omega&0\cr
0&E-\omega\cr}\right)\ ,
\eqno(2.3)
$$
where $E$ is the energy of the atoms in the incident beam. On the other
hand the splitting in energy $2\omega$ among the two internal
beams is usually induced by the action of laboratory controlled
effects, typically the presence of external fields.
For open system though, even in absence of external fields,
the quantity $\omega$ is in general nonvanishing. Indeed, one can
show that the weak interaction of the system with the external
environment induces in general a hamiltonian contribution, giving
rise to the ``Lamb shift'' term $\omega$ in (2.3) [1-3, 16, 29].

Nevertheless, it is the additional piece $D[\rho]$ in (2.2)
that describes true mixing enhancing phenomena: in absence
of it, the evolution of $\rho$ would be unitary and reversible.
It can be represented by a trace-preserving linear map acting
on the three independent components of the density matrix in (2.1).
Decomposing for convenience $\rho_3$ in its real and imaginary
parts,
$$
\rho_3=\rho^1-i\rho^2\ ,
\eqno(2.4a)
$$
and introducing the combination
$$
\rho_1-\rho_2=2\rho^3\ ,
\eqno(2.4b)
$$
one can then write $D[\rho]$ as a $3\times 3$ real,
symmetric matrix $\cal D$, acting on the real vector
$|\rho\rangle$ of components $(\rho^1,\rho^2,\rho^3)$:
$$
{\cal D}=-2\left[\matrix{a&b&c\cr
                           b&\alpha&\beta\cr
                           c&\beta&\gamma\cr}\right]\ .
\eqno(2.5)
$$
The six parameters $a$, $b$, $c$, $\alpha$, $\beta$ and $\gamma$,
with $a$, $\alpha$ and $\gamma$ non negative,
are not all independent; physical consistency of the full
time evolution ({\it i.e.} the request of complete positivity, see [38, 39]
for details) further imposes the following inequalities:
$$
\eqalign{
&2R\equiv\alpha+\gamma-a\geq0\ ,\cr
&2S\equiv a+\gamma-\alpha\geq0\ ,\cr
&2T\equiv a+\alpha-\gamma\geq0\ ,\cr
&RST\geq 2\, bc\beta+R\beta^2+S c^2+T b^2\ .
}\hskip -1cm
\eqalign{
&RS\geq b^2\ ,\cr
&RT\geq c^2\ ,\cr
&ST\geq\beta^2\ ,\cr
&\phantom{\beta^2}\cr
}\eqno(2.6)
$$

If one includes also the hamiltonian contribution and further recalls
that ${\rm Tr}[\rho(t)]=1$, the evolution equation (2.2) can be
rewritten as a diffusion equation for the 3-vector $|\rho(t)\rangle$:
$$
{\partial\over\partial t}\,|\rho(t)\rangle=
-2\,{\cal H}\, |\rho(t)\rangle\ ,\qquad
{\cal H}=\left[\matrix{a&b+\omega&c\cr
                           b-\omega&\alpha&\beta\cr
                           c&\beta&\gamma\cr}\right]\ .
\eqno(2.7)
$$
Its solution involves the exponentiation of the matrix $\cal H$:
$$
|\rho(t)\rangle={\cal M}(t)\, |\rho(0)\rangle\ ,\qquad
{\cal M}(t)=e^{-2{\cal H}t}\ ,
\eqno(2.8)
$$
where $|\rho(0)\rangle$ represents the initial state of the
atoms entering the interferometer. It coincides with one of the
following density matrices
$$
\rho^{(1)}={1\over2}\left(\matrix{
1&1\cr
1&1\cr
}\right)\ ,\qquad
\rho^{(2)}={1\over2}\left(\matrix{
\phantom{-}1&-1\cr
-1&\phantom{-}1\cr
}\right)\ ,
\eqno(2.9)
$$
corresponding to the possible choices of orientation of the
incident atomic beam with respect to the first diffracting
grating. Both choices lead to the same final results;
for definiteness, in the following we shall work with $\rho^{(1)}$,
so that $|\rho(0)\rangle=(1/2,0,0)$.

\vskip 1cm

{\bf 3. OBSERVABLES}
\medskip

In the language of density matrices, physical observables are
represented by suitable hermitian operators, whose mean values
can be obtained by taking their trace with $\rho(t)$.
In particular, the intensity pattern observed at the end
of the interferometer is given by the mean value of the
following projector operators [18, 27]:
$$
{\cal O}_+={1\over2}\left(\matrix{
1&e^{-i\theta}\cr
e^{i\theta}&1\cr
}\right)\ ,\qquad\quad
{\cal O}_-={1\over2}\left(\matrix{
1&e^{-i(\theta+\pi)}\cr
e^{i(\theta+\pi)}&1\cr
}\right)\ ,
\eqno(3.1)
$$
that correspond to the two possible exit beams in which
an atom can be found while exiting the apparatus.
In the standard situation, it is the phase $\theta$
that gives the modulation of the interferometric pattern.
This is usually obtained by moving the transverse position
of one of the gratings (or laser standing waves) responsible
for the diffraction of the atom beam. Indeed, in an idealized
situation one finds [30-32]:
$$
\theta=\kappa\, (x_1-2x_2+x_3)\ ,
\eqno(3.2)
$$
where $x_i$, $i=1,2,3$, represents the position, transverse
with respect to the incident beam, of the $i$-th grating,
while $\kappa$ is the wave vector of the diffracting lattice.

The intensity ${\cal I}_\pm$ of the interference figure
detected at the two possible exits is then given by:
$$
{\cal I}_\pm(t)=\langle{\cal O}_\pm\rangle
\equiv{\rm Tr}[{\cal O}_\pm\,\rho(t)]
={1\over 2}+{\cal O}_\pm^1\,\rho^1(t)+{\cal O}_\pm^2\,\rho^2(t)
+{\cal O}_\pm^3\,\rho^3(t)\ ,
\eqno(3.3)
$$
where definitions similar to the ones in (2.4) have been introduced
also for the entries of the two matrices ${\cal O}_\pm$.
Using (3.1), one finds:
$$
{\cal I}_\pm(t)={1\over2}\Big\{1\pm2\Big[\cos\theta\, \rho^1(t)
+\sin\theta\, \rho^2(t)\Big]\Big\}\ .
\eqno(3.4)
$$
Since an atom exiting the interferometer can only be found in
one of the two exit beams, particle conservation requires
${\cal I}_+(t)+{\cal I}_-(t)=1$, which is clearly satisfied
by (3.4).

The intensity curves in (3.4) can be compared with the
experiment, provided explicit expressions for the entries of
the matrix ${\cal M}(t)$ in (2.8) are given. Formally, this
can be obtained by studying the eigenvalue problem for
the $3\times3$ matrix $\cal H$ in (2.7):
$$
{\cal H}\, |v_k\rangle=\lambda_k\, |v_k\rangle\ ,\qquad
k=1,2,3\ .
\eqno(3.5)
$$
The three eigenvalues $\lambda_1$, $\lambda_2$, $\lambda_3$
satisfy a cubic equation,
$$
\lambda^3+r\,\lambda^2+s\,\lambda+w=\,0\ ,
\eqno(3.6)
$$
with real coefficients: $r=-(a+\alpha+\gamma)$,
$s=a\alpha+a\gamma+\alpha\gamma-b^2-c^2-\beta^2+\omega^2$,
$w=-{\rm det}{\cal H}$. It then follows that $\lambda_k$ are either
all real, or one is real and the remaining two are complex
conjugate; further, in both situations, one can show that
in presence of dissipation the three eigenvalues have always
positive real parts [40].

Using the fact that the matrix $\cal H$ itself obeys the
equation (3.6), one finds:
$$
\eqalign{
{\cal I}_\pm(t)={1\over2}\bigg\{1\pm \sum_{k=1}^3
{e^{-2\lambda_k\, t}\over 3\lambda_k^2 +2r\lambda_k + s}
\bigg[\Big(\lambda_k^2 -(\alpha+\gamma)\lambda_k+
\alpha\gamma-\beta^2\Big)\, &\cos\theta\cr
+\Big((\omega-b)(\lambda_k-\gamma)-\beta c\Big)\, &\sin\theta
\bigg]\bigg\}\ .}
\eqno(3.7)
$$
 From this general expression, one sees that in presence
of complex eigenvalues, a further harmonic
modulation in time of the interference figures occurs,
while exponential damping terms always prevail for
long enough times. Further, note that in absence of
dissipation, $a=b=c=\alpha=\beta=\gamma=\omega=\,0$,
and thus $\lambda_k=\,0$, the expressions of
${\cal I}_\pm$ in (3.7) reduce to their standard,
time-independent ones [30-32]:
$$
{\cal I}_\pm={1\over2}\big\{1\pm\cos\theta\big\}\ ;
\eqno(3.8)
$$
any deviation from this formula as described by (3.7)
clearly signals the presence of dissipative phenomena
in atom interferometry.

Although explicit expressions for the eigenvalues $\lambda_k$
can always be found via Cardano's formula [41], the form (3.7)
of the intensities ${\cal I}_\pm$ is rather involved,
and of limited use in practice. Having in mind possible
comparison with experimental data, the study of suitable
approximations of (3.7) might result appropriate.

In this respect, a useful working assumption is to take
$\gamma=\,0$;%
\footnote{$^\dagger$}{There are essentially two known ways
of implementing the condition of weak interaction between subsystem
and environment [1-3]: the singular coupling
limit (in which the time-correlations in the environment are assumed to 
be much
smaller than the typical time scale of the subsystem)
and the weak coupling limit (in which it is the subsystem
characteristic time scale that becomes large).
One can check that the second situation
leads precisely to the condition $\gamma=\,0$ [29].}
in this case, the inequalities (2.6) further impose
$b=c=\beta=\,0$ and $a=\alpha$. In this simplified
situation, the formula in (3.7) reduces to:
$$
{\cal I}_\pm(t)={1\over2}\Big\{1\pm e^{-2\alpha t}\,
\cos\big(\theta-2\omega t\big)\Big\}\ .
\eqno(3.9)
$$
This is surely the most simple expression that the intensity
probabilities ${\cal I}_\pm(t)$ take in presence of dissipative
effects. It differs from the standard expression in (3.8)
by the presence of an exponential damping factor and of
an additional harmonic phase that accumulates in time.

A different approximation of the general formula (3.7)
can be obtained when the parameters $a$, $b$, $c$, $\alpha$,
$\beta$ and $\gamma$ are small with respect of $\omega$.
This could happen when the interferometer is immersed in
a strong external field, so that its contribution to
the energy shift $\omega$ due to its interaction with the
atom beams largely overrides the one coming from the effects
of a weakly coupled environment.%
\footnote{$^\ddagger$}{In this respect, it should be noted that
even in absence of any external field, a hierarchy between $\omega$
and the other dissipative parameters $a$, $b$, $c$, $\alpha$,
$\beta$ and $\gamma$ could be nevertheless generated by the interaction
with the environment. For details, see [16] and the Appendix in [29].}
In this case, the additional piece $D[\rho]$ in the evolution
equation (2.2) can be treated as a perturbation. Using the solution
of this equation expanded up to second order in the small
parameters, from (3.4) one obtains:
$$
\eqalign{
{\cal I}_\pm(t)={1\over2}\bigg\{1\pm e^{-(a+\alpha)t} \bigg[\bigg(
\cos2\Omega t &+{\alpha-a\over 2\Omega}\sin2\Omega t
-{2\beta^2\over\Omega^2}\sin^2\Omega t\bigg)\, \cos\theta\cr
&+\bigg({b-\omega\over\Omega}\sin2\Omega t-
{c\beta\over\Omega^2}\cos2\Omega t\bigg)\, \sin\theta
\bigg]\bigg\}\ ,}
\eqno(3.10)
$$
where $\Omega=\big[\omega^2-b^2-c^2-\beta^2-(\alpha-a)^2/4]^{1/2}$.
In writing (3.10) we have reconstructed the exponential factor
by putting together the terms linear and quadratic in $t$;
a similar treatment has allowed writing all harmonic pieces
in terms of the frequency $\Omega$. It is worth noting that
for $c=\beta=\,0$ the formula (3.10) gives the exact expression
for the intensities ${\cal I}_\pm$: no approximation is
involved. This is a consequence of the fact that for
$c=\beta=\,0$ the matrix $\cal H$ in (2.7) becomes block diagonal,
so that explicit, manageable expressions for its exponential
${\cal M}(t)$, and therefore for ${\cal I}_\pm$, can
be given. From this point of view, the validity of (3.10) goes
beyond the second order approximation in which it has been
derived: it can be considered as the expansion of the full
expression (3.7) for ${\cal I}_\pm$ up to second order in
$c$ and $\beta$.

\vskip 1cm

{\bf 4. INTERFERENCE PATTERN}
\medskip

The behaviour of the general expression (3.7) for ${\cal I}_\pm$
and of its special cases (3.9) and (3.10) crucially depend on the
dissipative parameters: at least in principle, they can be used
to obtain informations on their values 
from fits with the experimental data.
The magnitude of $a$, $b$, $c$, $\alpha$,
$\beta$, $\gamma$ and $\omega$ are nevertheless expected to be
very small. In fact, for subsystems in interactions with large
environments, the effects leading to dissipation and
decoherence can be roughly estimated to be proportional
to the typical energy of the system, while suppressed by
inverse powers of the characteristic energy scale of the
environment [1-3]. In the case of non-standard phenomena
induced by the dynamics of fundamental objects (strings, branes)
at Planck's mass $M_P$, an upper bound on the magnitude of
the dissipative parameters can be roughly evaluated to be
of order $M_A^2/M_P$, where $M_A$ is the mass of the atoms
used in the interferometer [16, 29]; in typical real situations,
this ratio takes values between $10^{-18}$ and $10^{-15}$ GeV
(or equivalently, between $10^3$ and $10^6$ KHz).

A further difficulty in comparing the theoretically predicted
interference figures with the experimental data arises from the
fact that the previously derived expressions for
${\cal I}_\pm$ hold in the case of an idealized interferometer,
with perfectly monoenergetic atomic incident beams.
In practice, the values of the atom momenta spread over a finite
distribution. This fact, together with the inevitable imperfections
in the construction of the actual interferometric apparatus,
produce attenuation in the intensity of the signal.

One can take into account these spurious effects by modifying
the previously derived intensity spectra with the introduction
of suitable unknown parameters. To keep the discussion as simple
as possible, we shall concentrate on the expression (3.9)
for ${\cal I}_\pm$; similar arguments apply to the other
formulas. By denoting with ${\cal N}_\pm$ the atom countings
at the two exit beams of the interferometer, one generalizes
the spectra in (3.9) as:
$$
{\cal N}_\pm(t)={\cal N}^{(0)}_\pm\Big\{1\pm {\cal C}_\pm\,
e^{-2\alpha t}\, \cos\big(\theta-2\omega t\big)\Big\}\ .
\eqno(4.1)
$$
The constants ${\cal C}_\pm$ are the fringe contrast and parametrize
the intensity attenuation, while ${\cal N}^{(0)}_\pm$ are suitable
normalization factors;%
\footnote{$^\dagger$}{In absence of dissipative effects, a theoretical
estimate of ${\cal C}_\pm$ has been obtained using atom optics [42, 43].}
note that particle conservation now requires:
${\cal N}^{(0)}_+\,{\cal C}_+={\cal N}^{(0)}_-\,{\cal C}_-$.
Clearly, the higher the fringe contrast, the more accurate the
determination of the dissipative parameters $\alpha$ and $\omega$
from the experiment will be.

In order to fit actual experimental data with the expression (4.1),
further elaborations are however needed. As mentioned before,
the intensity spectra are reconstructed by counting the atoms
at one of the exit beams as a function of the transverse position
$x$ of the final grating (or standing laser wave), with respect
to a reference, initial situation. This means that the geometry
of the two paths followed by the atoms inside the interferometer
slightly changes as $x$ varies;%
\footnote{$^\ddagger$}{The situation is completely different
in a neutron interferometer [34]: made of a silicon crystal,
its geometry can not be varied. In this case, the interferometric
spectra are obtained through a thin slab of material inserted
transversally to the two beams inside the interferometer;
a slight rotation of it produces a phase difference $\omega$
between the two ``optical'' paths.}
as a consequence, also the total evolution time $t$, the time
spent by the atoms inside the interferometer, changes with $x$.
Since the two path inside the
apparatus are very close to each other (their actual separation is
at most 100 $\mu$m), for small $x$ one finds:
$$
t=t_0+{\vartheta_1\over v}x\ ,
\eqno(4.2)
$$
where $v$ is the average velocity of the atoms in the incident
beam, while $\vartheta_1$ is the first order Bragg diffraction
angle (typically of order $10^{-4}$ rad). On the other hand
$t_0$ is the fixed time of flight of the atoms inside the
interferometer when it is in its initial, reference status;
it can be determined with high accuracy from the geometric
specifications of the actual apparatus and can be modified
only by changing the energy of the primary atom beam,
or by modifying the longitudinal dimension of the interferometer.

The outcome of this discussion is that
it should be possible to estimate the values of
$\alpha$ and $\omega$ from the behaviour in $x$ of the expression
in (4.1), and therefore from a fit with experimental data.
Unfortunately, in the experimental set-ups so far constructed
the dependence of $t$ on $x$ can hardly be seen: one finds
that while $t_0$ is at most $10^{-3}$ sec, the quantity
$\vartheta_1 x/v$ results at least ten orders of magnitude
smaller, even for maximal values of the displacement $x$
(a few hundreds nm).

Therefore, as a good approximation one can safely take
$t\simeq t_0$, and rewrite the interference pattern (4.1) as
$$
{\cal N}_\pm(x)={\cal N}^{(0)}_\pm\Big\{1 \pm\Big[
{\cal P}_\pm\,\cos(\theta_0+\kappa x)
+ {\cal Q}_\pm\,\sin(\theta_0+\kappa x)\Big]\Big\}\ ,
\eqno(4.3)
$$
where
$$
{\cal P}_\pm={\cal C}_\pm\, e^{-2\alpha t_0}\, \cos2\omega t_0\ ,\qquad
{\cal Q}_\pm={\cal C}_\pm\, e^{-2\alpha t_0}\, \sin2\omega t_0\ ,
\eqno(4.4)
$$
while $\theta_0$ is a fixed phase that is characteristic of each
interferometer. A fit of (4.3) with experimental data will allow to
determine the parameters ${\cal N}^{(0)}_\pm$, ${\cal P}_\pm$,
${\cal Q}_\pm$ and $\theta_0$, and therefore to obtain informations
on the dissipative parameters.

\vskip 1cm

{\bf 5. DISCUSSION}
\medskip

We shall now briefly report on the results of a 
$\chi^2$ fit of the formula (4.3) with recently published
data from two experiments [44, 45].
The analysis that follows is of limited quantitative meaning:
direct access to the data and a careful study of systematic errors 
are needed in order to obtain precise determination of the dissipative
effects; nevertheless, it will provide a rough estimate about
the sensitivity of present atom interferometers
to the parameters $\alpha$ and $\omega$.

The atom interferometers used in the two
experiments are particularly sensitive devices, reaching a very high fringe
contrast; this is obtained by using neon,
respectively lithium, atoms as ``matter waves'' and
laser standing light as diffracting device.
Both in [44] and [45], only data from one of the two exit beams
are reported, the ones corresponding to the lower sign in (4.3).
Since as mentioned before $t_0$ is known with high precision,
from the ratio ${\cal P}_-/{\cal Q}_-$
one can immediately obtain an estimate for the parameter $\omega$.
Using the data from the first experiment, from our fit we find
$\omega=(0.7\pm 0.2)\times 10^{-21}\ {\rm GeV}$, where the quoted
error is only statistical.

On the other hand, the determination of $\alpha$ is subordinated to
the estimate of the fringe contrast ${\cal C}_-$. This would not
be necessary if the parameters ${\cal P}_-$ and ${\cal Q}_-$ can
be measured for two different values of the flight time $t_0$.
As mentioned before, this can be obtained either by changing the
average velocity of the incoming atoms, or by varying the dimensions
of the interferometer. In lacking of this extra information,
we shall obtain an estimate for ${\cal C}_-$ using directly the data.

In absence of dissipative effects, $\alpha=\omega=\,0$, the constant
${\cal C}_-$ can be obtained from the maximum, ${\cal N}^{(\rm max)}$,
and the minimum, ${\cal N}^{(\rm min)}$, atom counts in the experimental
interference figure:
${\cal C}_-=({\cal N}^{(\rm max)}-{\cal N}^{(\rm min)})/
({\cal N}^{(\rm max)}+{\cal N}^{(\rm min)})$.
Although this formula is only approximately valid for nonvanishing 
$\alpha$
and $\omega$, in practice the systematic error that one makes
in adopting it can be estimated to be at the end much smaller
than the pure experimental uncertainty. Using the rough
experimental data, one then deduces: ${\cal C}_-\simeq 62\%$.
With this value, one finally gets:
$\alpha=(0.1\pm 0.1)\times 10^{-22}\ {\rm GeV}$, which is compatible
with zero.

The accuracy in the determination of $\alpha$ and $\omega$ improves
using the data from the most recent experiment, thanks to the
higher fringe contrast (of about $74\%$) and the increase in
the number of experimental points. In fact, the same procedure
adopted before now gives the following estimates:
$\alpha=(0.3 \pm 0.1)\times 10^{-23}\ {\rm GeV}$ and
$\omega=(0.20 \pm 0.01)\times 10^{-21}\ {\rm GeV}$.~\hskip -.1cm
\footnote{$^\dagger$}{We remark that as before the quoted 
errors are purely statistical; a thorough analysis of the
full experimental data, that takes into account also the
systematics, would likely worsen the estimated errors,
in particular that on $\omega$.}
Note that these values are perfectly compatible with the
ones previously determined in the case of the neon
interferometer. As explained before, the values
of the dissipative parameters should be proportional to
the square of the mass of the atoms in the incident beams.
Therefore, the values of $\alpha$ and $\omega$ determined
with the data from the lithium beam should result smaller
than those obtained from the first experiment.

In conclusion, the results of our discussion
show that atom interferometers are potentially very
sensitive to the presence of phenomena leading to dissipation
and decoherence. Although the performed error analysis has
been limited to statistical uncertainties, the derived estimates
seem to indicate nonvanishing values for $\alpha$ and
$\omega$, of magnitude compatible with 
an origin from a fundamental dynamics at a very large mass scale. 
As already remarked, direct access to the rough experimental
data and more complete $\chi^2$ fits are needed
in order to claim the presence of dissipative effects.
We nevertheless hope that our preliminary analysis will
stimulate further, more accurate investigations.

\vfill\eject

\centerline{\bf REFERENCES}
\bigskip\medskip

\item{1.} R. Alicki and K. Lendi, {\it Quantum Dynamical Semigroups and
Applications}, Lect. Notes Phys. {\bf 286}, (Springer-Verlag, Berlin, 
1987)
\smallskip
\item{2.} V. Gorini, A. Frigerio, M. Verri, A. Kossakowski and
E.C.G. Surdarshan, Rep. Math. Phys. {\bf 13}, 149 (1978)
\smallskip
\item{3.} H. Spohn, Rev. Mod. Phys. {\bf 53}, 569 (1980)
\smallskip
\item{4.} A. Royer, Phys. Rev. Lett. {\bf 77}, 3272 (1996)
\smallskip
\item{5.} G. Lindblad, Comm. Math. Phys. {\bf 48}, 119 (1976)
\smallskip
\item{6.} W.H. Louisell, {\it Quantum Statistical Properties of 
Radiation},
(Wiley, New York, 1973)
\smallskip
\item{7.} M.O. Scully and M.S. Zubairy,
{\it Quantum Optics} (Cambridge University Press, Cambridge, 1997)
\smallskip
\item{8.} C.W. Gardiner and P. Zoller,
{\it Quantum Noise}, 2nd. ed. (Springer, Berlin, 2000)
\smallskip
\item{9.} L. Fonda, G.C. Ghirardi and A. Rimini, Rep. Prog. Phys.
{\bf 41}, 587 (1978)
\smallskip
\item{10.} H. Nakazato, M. Namiki and S. Pascazio,
Int. J. Mod. Phys. {\bf B10}, 247 (1996)
\smallskip
\item{11.} F. Benatti and R. Floreanini, Phys. Lett. {\bf B428}, 149 
(1998) 
\smallskip
\item{12.} D. Giulini, E. Joos, C. Kiefer, J. Kupsch, I.-O. Stamatescu
and H.D. Zeh, {\it Decoherence and the Appearance of a Classical World 
in Quantum Theory}, (Springer-Verlag, Berlin, 1996)
\smallskip
\item{13.} S. Sinha, Phys. Lett. {\bf A228} (1997) 1
\smallskip
\item{14.} G.C. Ghirardi, A. Rimini and T. Weber, Phys. Rev. D {\bf 34}, 
470 (1986)
\smallskip
\item{15.} J. Ellis, N.E. Mavromatos and D.V. Nanopoulos, Phys. Lett.
{\bf B293}, 37 (1992); Int. J. Mod. Phys. {\bf A11}, 1489 (1996)
\smallskip
\item{16.} F. Benatti and R. Floreanini, Ann. of Phys. {\bf 273},
58 (1999) 
\smallskip
\item{17.} S. Hawking, Comm. Math. Phys. {\bf 87}, 395 (1983); Phys. 
Rev. D {\bf 37}, 904 (1988); Phys. Rev. D {\bf 53}, 3099 (1996);
S. Hawking and C. Hunter, Phys. Rev. D {\bf 59}, 044025 (1999)
\smallskip
\item{18.} J. Ellis, J.S. Hagelin, D.V. Nanopoulos and M. Srednicki,
Nucl. Phys. {\bf B241}, 381 (1984)
\smallskip
\item{19.} S. Coleman, Nucl. Phys. {\bf B307}, 867 (1988)
\smallskip
\item{20.} S.B. Giddings and A. Strominger, Nucl. Phys. {\bf B307}, 
854 (1988)
\smallskip
\item{21.} M. Srednicki, Nucl. Phys. {\bf B410}, 143 (1993)
\smallskip
\item{22.} W.G. Unruh and R.M. Wald, Phys. Rev. D {\bf 52}, 2176 (1995)
\smallskip
\item{23.} L.J. Garay, Phys. Rev. Lett. {\bf 80}, 2508 (1998);
Phys. Rev. D {\bf 58}, 124015 (1998)
\smallskip
\item{24.} For a presentation of these models, see: V.A. Rubakov,
Russian J. Phys. {\bf 44}, 871 (2001)
\smallskip
\item{25.} F. Benatti and R. Floreanini, Nucl. Phys. {\bf B488},
335 (1997) 
\smallskip
\item{26.} F. Benatti, R. Floreanini and R. Romano,
Nucl. Phys. {\bf B602}, 541 (2001)
\smallskip
\item{27.} F. Benatti and R. Floreanini, Phys. Lett. {\bf B451},
422 (1999) 
\smallskip
\item{28.} F. Benatti and R. Floreanini, Phys. Rev. D {\bf 62},
125009 (2000) 
\smallskip
\item{29.} F. Benatti and R. Floreanini, Phys. Rev. D {\bf 64},
085015 (2001)
\smallskip
\item{30.} {\it Atom Interferometry}, P.R. Berman, ed., (Academic Press,
New York, 1997)
\smallskip
\item{31.} J. Baudon, R. Mathevet and J. Robert, J. Phys. B
{\bf 32}, R173 (1999)
\smallskip
\item{32.} R.M. Godun, M.B. D'Arcy, G.S. Summy and K. Burnett,
Contemp. Phys. {\bf 42}, 77 (2001)
\smallskip
\item{33.} P. Meystre, {\it Atom Optics}, (Springer-Verlag, New York, 
2001)
\smallskip
\item{34.} H. Rauch and S.A. Werner, {\it Neutron Interferometry}
(Oxford University Press, Oxford, 2000)
\smallskip
\item{35.} B.-G. Englert, C. Miniatura and J. Baudon,
J. Phys. II France {\bf 4}, 2043 (1994)
\smallskip
\item{36.} R.A. Rubenstein {\it et al.}, Phys. Rev. Lett. {\bf 82},
2018 (1999)
\smallskip
\item{37.} D.A. Kokorowski, A.D. Cronin, T.D. Roberts and D.E. Pritchard,
Phys. Rev. Lett. {\bf 86}, 2191 (2001)
\smallskip
\item{38.} V. Gorini, A. Kossakowski and E.C.G. Sudarshan,
J. Math. Phys. {\bf 17}, 821 (1976)
\smallskip
\item{39.} F. Benatti and R. Floreanini,
Mod. Phys. Lett. {\bf A12}, 1465 (1997);
Phys. Lett. {\bf B468}, 287 (1999)
\smallskip
\item{40.} K. Lendi, J. Phys. {\bf A 20}, 13 (1987)
\smallskip
\item{41.} M. Artin, {\it Algebra}, (Prentice Hall, Englewood Cliffs,
1991)
\smallskip
\item{42.} C. Champenois, M. B\"uchner and J. Vigu\'e,
Eur. Phys. J. D {\bf 5} (1999) 363
\smallskip
\item{43.} R. Delhuille {\it et al.}, Comptes  Rendus Acad. Sci. Paris,
Series IV {\bf 2}, 587 (2001)
\smallskip
\item{44.} D.M. Giltner, R.W. McGowan and S. Au Lee, Phys. Rev. Lett.
{\bf 75}, 2638 (1995)
\smallskip
\item{45.} R. Delhuille, C. Champenois, M. B\"uchner
L. Jozefowski, C. Rizzo, G. Tr\'enec and J. Vigu\'e,
High contrast Mach-Zender lithium atom interferometer in the
Bragg regime, {\tt quant-ph/0111052}, 2001

\bye